\documentstyle [12pt]{article}
\renewcommand{\theequation}{\arabic{equation}}
\begin{document}
\hfill UTHEP--93--1001\vskip.01truein
\hfill{Oct., 1993}\vskip1truein
\centerline{\Large Multiple Gluon Effects in Fermion--(Anti)Fermion}
\centerline{\Large Scattering at SSC/LHC
            Energies{\let\thefootnote=\ast\footnote{
Research supported in part by the Texas National Research Laboratory Commission
under contracts RCFY9101, RCFY9201, and RCFY--93--347 and by Polish Government
grants KBN 20389101 and KBN 223729102.}}}
\vskip.8truein
\centerline{\sc D.~B.~DeLaney, S.~Jadach,{\let\thefootnote=\dag\footnote{
Permanent address: Institute of Nuclear Physics, ul. Kawiory 26a, Cracow,
Poland.}} Ch.~Shio, G.~Siopsis, and B.~F.~L.~Ward}\vskip.5truein
\centerline{\it Department of Physics and Astronomy}
\centerline{\it The University of Tennessee, Knoxville, TN 37996--1200}
\centerline{\it U. S. A.}\vskip.05truein\baselineskip=21pt\vskip.6truein
\centerline{\bf ABSTRACT}\vskip.2truein\par
\let\sstl=\scriptscriptstyle
We extend the methods of Yennie, Frautschi and Suura (YFS) to compute, via
Monte
Carlo methods, the effects of multiple gluon emission in the processes
$q+{\llap
{\phantom q}^{\sstl(}\bar q^{\sstl)}{}}'\to q''+{\llap{\phantom q}^{\sstl(}\bar
q^{\sstl)}{}}'''+n(G)$, where $G$ is a soft gluon. We show explicitly that the
infrared singularities in the respective simulations are canceled to
all orders
in $\alpha_s$. Some discussion of this result from the standpoint of
confinement
is given. More importantly, we present, for the first time ever, sample
numerical Monte Carlo data on multiple soft gluon emission in the rigorously
extended YFS framework. We find that such soft gluon effects must be taken into
account for precise SSC/LHC physics simulations.
\par\renewcommand\thepage{}
\vfill\eject\parskip.1truein\parindent=20pt\pagenumbering{arabic}\par
\section{Introduction}\label{intro}\par
While the SSC has been cancelled and is being re-evaluated, the LHC
planning stages continue to gain momentum, so that
it still becomes more and more necessary to prepare for the physics
exploration which the LHC, at the least, will provide.
The primary issue in
this exploration is the comparison between predictions from the standard
$SU_{2L
}\otimes U_1\otimes SU_3^C$ theory and its possible extensions, and what will
be
observed by the GEM and SDC collaborations, with the discovery of the Higgs
particle or 
whatever it represents as of course a primary goal of such comparisons. Thus,
it
is important to know the theoretical predictions as accurately as is necessary
to exploit the expected detector performances of GEM and SDC to the fullest
extent. In particular, higher order radiative corrections to these predictions
(due to multiple photon and multiple gluon effects) are in fact essential to
obtain the proper precision on the signal and background processes in the
respective SSC/LHC environment. In Refs.~\cite{dbd}, we have developed
Yennie--Frautschi--Suura \cite{yfs} (YFS) multiphoton Monte Carlo event
generators for calculating $n\gamma$ effects in SSC/LHC processes.
In what follows,
we now develop the extension of our own YFS Monte Carlo--based higher order
radiative correction methods to multiple gluon effects in SSC/LHC processes.
Such a
calculation of $n(G)$ effects has not appeared elsewhere, where we use $G$ to
denote a gluon.\par
More precisely, we want to use the fact that, by the uncertainty principle,
infinite--wavelength gluons should not affect the motion of quarks and gluons
inside a proton, whose radius is $\sim1$ fm. Thus, we have recently realized
\cite{preprint} this physical requirement in perturbation theory by showing
that
in our prototypical processes $q+{\llap{\phantom q}^{\sstl(}\bar q^{\sstl)}{}}'
\to q''+{\llap{\phantom q}^{\sstl(}\bar q^{\sstl)}{}}'''+(G)$, the infrared
singularities cancel at $O(\alpha_s)$, just as they cancel in the analogous QED
process $f+f'\to f+f'+\gamma$. We then are able to define the QCD analogues of
the famous YFS infrared functions \cite{yfs} B and \~B which describe the
virtual and real infrared singularities in QED processes to all orders in
$\alpha$. These functions, $B_{\rm QCD}$ and $\tilde B_{\rm QCD}$, then
describe
the infrared singularities in QCD in the processes $q+{\llap{\phantom
q}^{\sstl(
}\bar q^{\sstl)}{}}'\to q''+{\llap{\phantom q}^{\sstl(}\bar
q^{\sstl)}{}}'''+n(G
)$ to all orders in $\alpha_S$. The sum $B_{\rm QCD}+\tilde B_{\rm QCD}$ is
infrared finite and allows us to extend our YFS Monte Carlo methods, widely
used
in SLC/LEP physics for higher order QED radiative effects, to the analogous
Monte Carlo methods in QCD processes for SSC/LHC physics objectives.
We call the
resulting programs SSCYFSG and SSCBHLG, for they are the QCD extensions of the
QED $n\gamma$ Monte Carlos SSCYFS2 and SSCBHL. Here we recall that \break
SSCYFS2 was already described in Ref.~\cite{dbd} and SSCBHL is the analogous
SSC
extension of the YFS Monte Carlo BHLUMI 2.01 \cite{jwwas} for the SLC/LEP
luminosity process $e^+e^-\to e^+e^-+n\gamma$.\par
In what follows, we shall present sample Monte Carlo data for the process $q+{
\llap{\phantom q}^{\sstl(}\bar q^{\sstl)}{}}'\to q''+{\llap{\phantom
q}^{\sstl(}
\bar q^{\sstl)}{}}'''+n(G)$ from the event generator SSCBHLG. Analogous results
from SSCYFSG will appear elsewhere \cite{toapp}. The latter Monte Carlo only
simulates initial--state gluon radiation, whereas SSCBHLG simulates
initial--state and final--state $n$--gluon radiation, as well as
the respective initial--final--state
interference effects. Further, we emphasize that (unlike BHLUMI
2.01) SSCBHLG is not restricted to small scattering angles. Note finally that
SSCYFSG is an initial--state restriction of SSCBHLG and hence is useful for
cross--checking our work.\par
Our work is organized as follows. In the next Section, we derive the extension
of the YFS exponentiation to $n$--gluon emission and the attendant extension of
SSCYFS2 and SSCBHL. In Section III, we present sample Monte Carlo data for
$n$--gluon emission for the SDC and GEM acceptances. Section IV contains some
concluding remarks.\par
\section{Infrared Singularity Cancellation in QCD}\label{cancel}\par
In this Section, we derive the analogue of the YFS infrared functions B and \~B
for QCD, \break$B_{\rm QCD}$ and $\tilde B_{\rm QCD}$, for the process $q+{
\llap{\phantom q}^{\sstl(}\bar q^{\sstl)}{}}'\to q''+{\llap{\phantom
q}^{\sstl(}
\bar q^{\sstl)}{}}'''+(G)$. We will focus on the case of most interest to us,
in
which a gluon is exchanged in the $t$--channel and we require that the
exchanged
gluon carries large momentum transfer so that the outgoing quark (anti--quark)
is in the $|\eta|\le2.8$ acceptance of the SDC (and GEM). This situation is
depicted in Fig.~\ref{fone}.
We remark that if the relevant
exchange is a color singlet exchange (such as $W$ or $Z^0$ exchange), we can
obtain the corresponding formulas for $B_{\rm QCD}$ and $\tilde B_{\rm QCD}$ by
dropping the terms in the color--exchange case which arise from
non--commutativity of the respective QCD $\lambda^a$ matrices in the relevant
quark representation.\par
Specifically, following the kinematics in Fig.~\ref{fone}a, we note that, for
the SDC and GEM acceptance, we always have $-Q^2=-(q_1-q_2)^2\gg\Lambda^2_{\rm
QCD}$, so that perturbative QCD methods are applicable.
What we wish to do is to
extract the infrared--singular part of the $O(\alpha_s)$ amplitude in
Fig.~\ref{fone}b, in complete analogy with the YFS \cite{yfs} extraction of the
infrared--singular portion of the analogous amplitude in QED via the famous YFS
virtual infrared function B. To this end, we first note that, by explicit
calculation, the graphs (v)-(vii) are not infrared--divergent (we compute the
gluon vacuum polarization to order $\alpha_s$ as well), so that the
infrared--singular part of Fig.~\ref{fone}b is given by just the same graphs as
the infrared--singular part of the analogous process in QED. We follow the YFS
methods in Ref.~\cite{yfs} and find the infrared--singular part of
Fig.~\ref{fone}b to be given by \begin{equation}{\cal M}_{\rm v,IR}=\alpha_s
B_{\rm QCD}{\cal M}_{8B}+\alpha_s\bar B_{\rm QCD}{\cal
M}_{0B}\quad,\label{eone}
\end{equation} where ${\cal M}_{8B}$ is the respective Born amplitude
in Fig.~\ref{fone}a and where
\let\ttl=\textstyle
\[B_{\rm QCD}={i\over(8\pi^3)}\int{d^4k\over(k^2-\lambda^2+i\epsilon)}\left[C_F
\left({\ttl2p_1+k\over\ttl k^2+2k\cdot p_1+i\epsilon}+{\ttl2p_2-k\over\ttl
k^2-2
k\cdot p_2+i\epsilon}\right)^2\right.\]\vskip-30pt\begin{eqnarray*}+\Delta C_s{
\ttl2(2p_1+k)\cdot(2p_2-k)\over\ttl(k^2+2k\cdot p_1+i\epsilon)(k^2-2k\cdot
p_2+i
\epsilon)}&+C_F&\left({\ttl2q_1+k\over\ttl k^2+2k\cdot q_1+i\epsilon}-{\ttl
2q_2
-k\over k^2-2k\cdot q_2+i\epsilon}\right)^2\\ +\Delta C_s{\ttl2(2q_1+k)\cdot(2
q_2-k)\over\ttl(k^2+2k\cdot q_1+i\epsilon)(k^2-2k\cdot q_2+i\epsilon)}&+C_F&
\left({\ttl2p_2+k\over\ttl k^2+2k\cdot p_2+i\epsilon}-{\ttl2q_2+k\over\ttl
k^2+2
k\cdot q_2+i\epsilon}\right)^2\\ +\Delta
C_t{\ttl2(2q_2+k)\cdot(2p_2+k)\over\ttl
(k^2+2k\cdot q_2+i\epsilon)(k^2+2k\cdot p_2+i\epsilon)}&+C_F&\left({\ttl2p_1+k
\over\ttl k^2+2k\cdot p_1+i\epsilon}-{\ttl2q_1+k\over\ttl k^2+2k\cdot q_1+i
\epsilon}\right)^2\\ +\Delta
C_t{\ttl2(2q_1+k)\cdot(2p_1+k)\over\ttl(k^2+2k\cdot
q_1+i\epsilon)(k^2+2k\cdot p_1+i\epsilon)}&-C_F&\left({\ttl2p_1+k\over\ttl
k^2+2
k\cdot p_1+i\epsilon}-{\ttl2q_2+k\over k^2+2k\cdot q_2+i\epsilon}\right)^2\\ +
\Delta C_u{\ttl2(2p_1+k)\cdot(2q_2+k)\over\ttl(k^2+2k\cdot
p_1+i\epsilon)(k^2+2k
\cdot q_2+i\epsilon)}&-C_F&\left({\ttl2q_1+k\over\ttl k^2+2k\cdot
q_1+i\epsilon}
-{\ttl2p_2+k\over\ttl k^2+2k\cdot p_2+i\epsilon}\right)^2\end{eqnarray*}
\begin{equation}+\Delta C_u{\ttl2(2q_1+k)\cdot(2p_2+k)\over\ttl(k^2+2k\cdot
q_1+
i\epsilon)(k^2+2k\cdot p_2+i\epsilon)}\label{etwo}\end{equation} with $C_F=4/3=
$quadratic Casimir invariant of the quark color representation, $\lambda$ equal
to the standard infrared regulator mass (Ref.~\cite{corn})
and \begin{eqnarray}\Delta C_s&=&\left\{\begin{array}{ll}-1,&qq'\mbox{
incoming}\\ -1/6,&q\bar q'\mbox{ incoming}\end{array}\right.,\quad\Delta
C_t=-3/
2,\mbox{ and}\nonumber\\ \Delta C_u&=&\left\{\begin{array}{ll}-5/2,&qq'\mbox{
incoming}\\ -5/3,&q\bar q'\mbox{ incoming}\end{array}\right..\label{ethree}
\end{eqnarray} Thus, we see that the non--Abelian nature of QCD causes $\Delta
C_j\ne0$.\par
The additional term in (\ref{eone}), which arises from color algebra, consists
of what we call the ``singlet'' exchange Born amplitude which is obtained from
Fig.~\ref{fone}a by substituting $\lambda^0$ for the color matrices $\lambda^a
$, where $\lambda^0\equiv I/\sqrt6$. The corresponding value of $\bar B_{\rm
QCD}$ in (\ref{eone}) is obtained from (\ref{etwo}) by setting $C_F\equiv0$ and
$\Delta C_t=0$ in (\ref{etwo}) and by setting\begin{equation}\Delta C_u=\Delta
C_s=\left\{\begin{array}{rl}-C_F,&qq'\mbox{ incoming}\\ C_F,&q\bar q'\mbox{
incoming}\end{array}\right.\label{efour}\end{equation} in (\ref{etwo}). We note
that, to order $\alpha_s$, the ``singlet'' amplitude does not contribute to our
cross section for the process in Fig.~\ref{fone}.\par
Turning now to the soft real emission process in Fig.~\ref{fone}b, we again
follow the methods of YFS in Ref.~\cite{yfs} and extract the
infrared--divergent
part of
the real gluon bremsstrahlung as\begin{equation}d\sigma^{B1}_{soft}=d\sigma_0(2
\alpha_s\tilde B_{\rm QCD})\label{efive}\end{equation} where the real QCD
infrared function is\begin{equation}2\alpha_s\tilde B_{\rm QCD}=\int^{k\le
K_{max}}{d^3k\over k_0}\tilde S_{\rm QCD}\!(k)\label{esix}\end{equation} with
\[\tilde S_{\rm QCD}\!(k)=-{\alpha_s\over4\pi}\left\{C_F\left({\ttl
p_1\over\ttl
p_1\cdot k}-{\ttl q_1\over\ttl q_1\cdot k}\right)^2+\Delta C_t{\ttl2p_1\cdot
q_1
\over\ttl k\cdot p_1k\cdot q_1}+C_F\left({\ttl p_2\over\ttl p_2\cdot k}-{\ttl
q_2\over\ttl q_2\cdot k}\right)^2+\Delta C_t{\ttl2p_2\cdot q_2\over\ttl k\cdot
p_2k\cdot q_2}\right.\]\[+C_F\left({\ttl p_1\over\ttl p_1\cdot k}-{\ttl
p_2\over
\ttl p_2\cdot k}\right)^2-\Delta C_s{\ttl2p_1\cdot p_2\over\ttl k\cdot
p_1k\cdot
p_2}+C_F\left({\ttl q_1\over\ttl q_1\cdot k}-{\ttl q_2\over\ttl q_2\cdot k}
\right)^2-\Delta C_s{\ttl2q_1\cdot q_2\over\ttl k\cdot q_1k\cdot q_2}\]
\vskip-.3in
\begin{equation}\left.-C_F\left({\ttl q_1\over\ttl q_1\cdot k}-{\ttl p_2\over
\ttl p_2\cdot k}\right)^2+\Delta C_u{\ttl2q_1\cdot p_2\over\ttl k\cdot
q_1k\cdot
p_2}-C_F\left({\ttl q_2\over\ttl q_2\cdot k}-{\ttl p_1\over\ttl p_1\cdot k}
\right)^2+\Delta C_u{\ttl2q_2\cdot p_1\over\ttl k\cdot q_2k\cdot p_1}\right\},
\label{eseven}\end{equation} where $K_{max}$ corresponds to the relevant gluon
jet detector
resolution energy ($\sim3$~GeV at the SSC/LHC) and $k_0\equiv\sqrt{\vec
k^2+\lambda^2}$ so that we regulate the infrared singularities in
(\ref{eseven})
in complete analogy with our gluon mass regulator in (\ref{etwo}). (Note that
graph (v) in Fig.~\ref{fone}c is not infrared--divergent.) Here, $d\sigma_0$ is
the respective Born differential cross section. The results (\ref{eone}) to
(\ref{eseven}) represent the complete infrared (IR) structure of QCD for the
cross section for $q+{\llap{\phantom q}^{\sstl(}\bar q^{\sstl)}{}}'\to q''+{
\llap{\phantom q}^{\sstl(}\bar q^{\sstl)}{}}'''+(G)$ to order $\alpha_s$. The
result (\ref{eseven}) is consistent with the pioneering work of Berends and
Giele in Ref.~\cite{bergi}. Following the YFS theory, we see that the criterion
for the cancellation of the $O(\alpha_s)$ and IR singularities in the cross
section for our scattering process in Fig.~\ref{fone} is that\begin{equation}{
\rm SUM_{IR}(QCD)}\equiv2\alpha_s\tilde B_{\rm QCD}+2\alpha_s{\rm Re}B_{\rm
QCD}
\label{eeight}\end{equation} should be independent of $\lambda$. Using the
well--known results \cite{yfs},\cite{ward} for the integrals in (\ref{etwo})
and
(\ref{eseven}) (notice that these are just the same integrals that arise in the
YFS analysis of QED with different color--based weights) we see that, indeed,
$\lambda$ does cancel out of (\ref{eeight}). We are left with the fundamental
result (for, e.g., $m_q=m_{q'}=m$) \begin{equation}{\rm
SUM_{IR}(QCD)}={\alpha_s
\over\pi}\sum_{A=\{s,t,u,s',t',u'\}}(-1)^{\rho(A)}(C_FB_{tot}(A)+\Delta
C_AB'_{t
ot}(A))\label{enine}\end{equation}
where\[B_{tot}(A)=\log(2K_{max}/\sqrt{|A|})^2
(\ln(|A|/m^2)-1)+{1\over2}\ln(|A|/m^2)-1-\pi^2/6+\theta(A)\pi^2/2\quad,\]
\begin{equation}B'_{tot}(A)=\log(2K_{max}/\sqrt{|A|})^2+{1\over2}\ln(|A|/m^2)-
\pi^2/6+\theta(A)\pi^2/2\quad \label{eten}\end{equation} and
\begin{equation}\rho(A)=\left\{\begin{array}{lrl}0,&A=s,s',&t,t'\\ 1,&A=u,u'&
\end{array}\right..\label{eleven}\end{equation}
The results
(\ref{enine}) and (\ref{eten}) are then the fundamental results of our
analysis.\par
Indeed, repeating the arguments in Ref.~\cite{yfs} and/or using the
factorization results in Ref.~\cite{bergi}, we see that by virtue of
(\ref{enine}), we may compute the hard gluon radiation residuals at
$O(\alpha_s)
$ as\begin{equation}\bar\beta_0=d\sigma^{1-loop}-2\alpha_s{\rm Re}B_{\rm QCD}d
\sigma_0\label{eeleven}\end{equation}
and\begin{equation}\bar\beta_1=d\sigma^{B1
}-\tilde S_{\rm QCD}d\sigma_0\label{etwelve}\end{equation}
in the exponentiated
formula \begin{equation}d\sigma_{exp}=exp[{\rm
SUM_{IR}(QCD)}]\sum_{n=0}^\infty{
1\over n!}\int\prod_{j=1}^n{d^3k_j\over k_j}\int{d^4y\over(2\pi)^4}e^{+iy(p_1+
p_2-q_1-q_2-\sum_jk_j)+D}\bar\beta_n(k_1,\dots,k_n){d^3q_1d^3q_2\over q_1^{\,0}
q_2^{\,0}}\label{ethirteen}\end{equation} where\begin{equation}D=\int{d^3k\over
k_0}\tilde S_{\rm QCD}\left[e^{-iy\cdot k}-\theta(K_{max}-|\vec
k|)\right]\quad.
\label{efourteen}
\end{equation}
Here, we use the virtual 1--loop correction
\begin{equation}
d\sigma^{1-loop}=d\sigma_0\delta_{vir}\,,
\label{efifteen}
\end{equation}
where
\begin{equation}\begin{array}{cl}
\delta_{vir}= {2\alpha_s\over\pi} & \left[ (C_F - {1\over 2} C_A) \left\{ \ln
{|t|\over m^2} \left(2-\ln {|t|\over \lambda^2} + {1\over 2} \ln {|t|\over m^2}
\right) \right. \right. \\ \\
& \quad\quad\quad\quad\quad\quad + \left. \ln {s\over m^2} \left(2 \ln {s\over
\lambda^2} -\ln {s\over m^2} \right)+ {tu\over s^2+u^2} \ln {s\over |t|}
\left({1\over 2} (s/u-1)\ln {s\over |t|} -1\right) \right\} \\ \\
& -(C_F - {1\over 4} C_A) \left\{ \ln {|u|\over m^2} \left(2\ln {|u|\over
\lambda^2} -  \ln {|u|\over m^2}\right) + {st\over s^2+u^2} \ln {u\over t}
\left({1\over 2} (1-u/s) \ \ln {u\over t} -1\right) \right\} \\ \\
& \left. + C_F \ln {|t|\over \lambda^2} + C_A \left(\ln {|t|\over m^2} +
1\right) + {31\over 36} C_A - {5\over 9} C_F + {\pi^2 \over 6} C_F \right] \\
\end{array}
\label{esixteen}
\end{equation}
The respective soft real gluon bremsstrahlung cross is given by
(\ref{efive}) with
\begin{equation}
\begin{array}{cl} 2\alpha_s \tilde{B}_{QCD} = {2\alpha_s\over\pi} & \left[ (C_F
- {1\over 2} C_A) \left\{ \ln {4K_{max}^2 \over \lambda^2} +\ln {|t|\over m^2}
\left( 1+ \ln {4K_{max}^2 \over \lambda^2} -{1\over 2} \ln {|t|\over m^2}
\right) \right. \right. \\ \\
& \quad\quad\quad\quad\quad\quad -2 \left. \ln {s\over m^2} \left( 1+ \ln
{4K_{max}^2 \over \lambda^2} -{1\over 2} \ln {s\over m^2} \right) \right\} \\
\\
& + \left. (C_F -{1\over 4} C_A) \left\{ -2 \ln {4K_{max}^2 \over
\lambda^2} + 2 \ln {|u|\over m^2} \left( 1+ \ln {4K_{max}^2 \over \lambda^2}
-{1\over 2} \ln {|u|\over m^2}\right) - {\pi^2\over 3} \right\} \right] \\
\end{array}
\label{eseventeen}
\end{equation}.
It allows $\bar\beta_0$ to be identified as
$d\sigma^{{\cal O}(\alpha_s)}_{soft}-2\alpha_s(\tilde B_{QCD}+B_{QCD})
d\sigma_0$, where $d\sigma^{{\cal O}(\alpha_s)}_{soft}=d\sigma^{1-loop}
+d\sigma^{B1}_{soft}$, for example.
The above expressions are for incoming $q\bar q$ and need to be modified
accordingly for $qq$ interactions. The hard gluon residual $\bar\beta_1$
makes a vanishing contribution in the soft gluon limit so that it will
be presented elsewhere \cite{toapp}; for, here we focus on the soft gluon
effects in (\ref{ethirteen}) so that we will not need $\bar \beta_1$
in the current discussion.

The results (\ref{eeleven}) and (\ref{efifteen}) -- (\ref{eseventeen}) are
now realized in (\ref{ethirteen})
via Monte Carlo methods by replacing the QED forms of $\bar
\beta_n$, $\rm SUM_{IR}$ as described in Refs.~\cite{dbd} for our SSCYFS2 and
SSCBHL event generators by the respective QCD forms. The resulting event
generators are SSCYFSG and SSCBHLG, where in the former, only only gluon
radiation from the initial state is treated. Sample Monte Carlo data from
SSCBHLG are illustrated in the next section. Similar data for SSCYFSG will
appear elsewhere \cite{toapp}.\par
We should stress that two limits of the QCD coupling constant are needed in
(\ref{eeleven}), (\ref{etwelve}) and (\ref{ethirteen}). One is the perturbative
$-q^2\gg\Lambda^2_{\rm QCD}$ regime in the hard gluon residuals $\bar\beta_n$
in
(\ref{ethirteen}). This regime is well--known and we use the standard type of
formula~\cite{gross}\begin{equation}\alpha_s(\mu)={\ttl12\pi\over\ttl(33-2n_f)
\ln(\mu^2/(\Lambda_{n_f}^{\overline{MS}})^2)}\label{eeighteen}\end{equation}
with $n_f=$number of open flavors and $\Lambda_n^{\overline{MS}}$ the
respective
value of $\Lambda_{\rm QCD}$: we take $\Lambda_4^{\overline{MS}}\simeq.238$~GeV
for definiteness.\par
The second regime occurs in SUM$_{\rm IR}$ in (\ref{enine}), where the
$k^2\to0$
limit is relevant for ``on--shell'' gluons of 4--momentum $k$. (We follow
Tarrah~\cite{tarrah} and use the concept of on--shell quarks and gluons in
perturbation theory at large momentum transfer.) Recently~\cite{mattingly}, it
has been pointed out by Mattingly and Stevenson that this limit of $\alpha_s/
\pi$ exists as an infrared fixed point and its limiting value is $\simeq.263$.
We use this value of $(\alpha_s/\pi)$ in SUM$_{\rm IR}$ in (\ref{enine}) and
(\ref{etwelve}). Because of the fixed point behavior, the value one uses for
the $k^2\to0$ coupling of gluons is not sensitive to the precise value
$\Lambda_4^{\overline{MS}}$, for example. Further, since we only work to
$O(\alpha_s)$ in the hard interactions, there is no contradiction between our
use of $\Lambda^{\overline{MS}}$ values in (\ref{eeighteen}) and the
renormalization scheme of Mattingly and Stevenson in Ref.~\cite{mattingly}.\par
We should also note that our light quark masses are always those determined by
Leutwyler and Gaisser in Ref.~\cite{gaesser}:
$m_u(1$~GeV$)\simeq5.1\times10^{-3
}$~GeV, $m_d(1$~GeV$)\simeq8.9\times10^{-3}$~GeV, $m_s(1$~GeV$)\simeq.175$~GeV,
$m_c(1$~GeV$)\simeq1.3$~GeV, $m_b(m_b)\simeq4.5$~GeV; for $m_t$, we take
$m_t(m_t)\simeq164$~GeV. The running of these masses is readily incorporated
into our calculations \cite{gaesser}; we ignore this running in the current
paper, since it does not affect our results at the level of accuracy of
interest
to us here.\par
With these explanatory remarks in mind, we now turn to explicit Monte Carlo
data
illustrations for multiple gluon effects in $q+{\llap{\phantom q}^{\sstl(}\bar
q^{\sstl)}{}}'\to q''+{\llap{\phantom q}^{\sstl(}\bar q^{\sstl)}{}}'''+n(G)$ at
SSC/LHC energies in the ATLAS-CMS(SDC-GEM) acceptances.\par
\section{Results}\label{results}\par
In this Section, we illustrate our multiple--gluon Monte Carlo event generators
in the sample case of $u+u\to u+u+n(G)$, where we require the out--going $u$
quarks to satisfy $|\eta|\le2.8$ and $E>58$~GeV, for definiteness, at $\sqrt s=
15.4$~TeV. (We comment about the SSC case $\sqrt s=40$~TeV as well.) For
completeness, we will show explicit Monte Carlo data for SSCBHLG, since in that
generator initial, initial--final interference, and final state $n(G)$
radiation
is simulated.\par
\begin{figure}
\setlength{\unitlength}{0.1mm}
\setlength{\unitlength}{0.1mm}
\begin{picture}(1600,1500)
\put(0,0){\framebox(1600,1500){}}
\put(300,250){\begin{picture}(1200,1200)
\put(0,0){\framebox(1200,1200){}}
\multiput(  240.00,0)(  240.00,0){   5}{\line(0,1){25}}
\multiput(     .00,0)(   24.00,0){  51}{\line(0,1){10}}
\multiput(  240.00,1200)(  240.00,0){   5}{\line(0,-1){25}}
\multiput(     .00,1200)(   24.00,  0){  51}{\line(0,-1){10}}
\put( 240,-25){\makebox(0,0)[t]{\large$  10.000$}}
\put( 480,-25){\makebox(0,0)[t]{\large$  20.000$}}
\put( 720,-25){\makebox(0,0)[t]{\large$  30.000$}}
\put( 960,-25){\makebox(0,0)[t]{\large$  40.000$}}
\put(1200,-25){\makebox(0,0)[t]{\large$  50.000$}}
\multiput(0,     .00)(0,  240.00){   6}{\line(1,0){25}}
\multiput(0,   24.00)(0,   24.00){  50}{\line(1,0){10}}
\multiput(1200,     .00)(0,  240.00){   6}{\line(-1,0){25}}
\multiput(1200,   24.00)(0,   24.00){  50}{\line(-1,0){10}}
\put(-25,   0){\makebox(0,0)[r]{\large$    .000\cdot10^{   3}$}}
\put(-25, 240){\makebox(0,0)[r]{\large$    .500\cdot10^{   3}$}}
\put(-25, 480){\makebox(0,0)[r]{\large$   1.000\cdot10^{   3}$}}
\put(-25, 720){\makebox(0,0)[r]{\large$   1.500\cdot10^{   3}$}}
\put(-25, 960){\makebox(0,0)[r]{\large$   2.000\cdot10^{   3}$}}
\put(-25,1200){\makebox(0,0)[r]{\large$   2.500\cdot10^{   3}$}}
\end{picture}}
\put(300,250){\begin{picture}(1200,1200)
\thinlines
\newcommand{\x}[3]{\put(#1,#2){\line(1,0){#3}}}
\newcommand{\y}[3]{\put(#1,#2){\line(0,1){#3}}}
\newcommand{\z}[3]{\put(#1,#2){\line(0,-1){#3}}}
\newcommand{\e}[3]{\put(#1,#2){\line(0,1){#3}}}
\y{   0}{   0}{   0}\x{   0}{   0}{  39}
\y{  39}{   0}{   0}\x{  39}{   0}{  40}
\y{  79}{   0}{   0}\x{  79}{   0}{  40}
\y{ 119}{   0}{   0}\x{ 119}{   0}{  40}
\y{ 159}{   0}{   0}\x{ 159}{   0}{  40}
\y{ 199}{   0}{   0}\x{ 199}{   0}{  40}
\y{ 239}{   0}{   0}\x{ 239}{   0}{  40}
\y{ 279}{   0}{   5}\x{ 279}{   5}{  40}
\y{ 319}{   5}{   2}\x{ 319}{   7}{  40}
\y{ 359}{   7}{  22}\x{ 359}{  29}{  40}
\y{ 399}{  29}{  51}\x{ 399}{  80}{  40}
\y{ 439}{  80}{   3}\x{ 439}{  83}{  40}
\y{ 479}{  83}{ 173}\x{ 479}{ 256}{  40}
\y{ 519}{ 256}{ 177}\x{ 519}{ 433}{  40}
\z{ 559}{ 433}{ 154}\x{ 559}{ 279}{  40}
\y{ 599}{ 279}{ 329}\x{ 599}{ 608}{  40}
\y{ 639}{ 608}{  64}\x{ 639}{ 672}{  40}
\z{ 679}{ 672}{ 308}\x{ 679}{ 364}{  40}
\y{ 719}{ 364}{ 259}\x{ 719}{ 623}{  40}
\z{ 759}{ 623}{ 121}\x{ 759}{ 502}{  40}
\z{ 799}{ 502}{ 304}\x{ 799}{ 198}{  40}
\y{ 839}{ 198}{  90}\x{ 839}{ 288}{  40}
\z{ 879}{ 288}{ 111}\x{ 879}{ 177}{  40}
\z{ 919}{ 177}{ 115}\x{ 919}{  62}{  40}
\y{ 959}{  62}{   6}\x{ 959}{  68}{  40}
\z{ 999}{  68}{  38}\x{ 999}{  30}{  40}
\z{1039}{  30}{  20}\x{1039}{  10}{  40}
\y{1079}{  10}{   0}\x{1079}{  10}{  40}
\z{1119}{  10}{   9}\x{1119}{   1}{  40}
\y{1159}{   1}{   0}\x{1159}{   1}{  40}
\end{picture}}
\end{picture}
\vskip 1in
\caption{Gluon multiplicity distribution in $u+u\to u+u+n(G)$ at $\protect\sqrt
s=15.4$~TeV.}
\label{ftwo}
\end{figure}
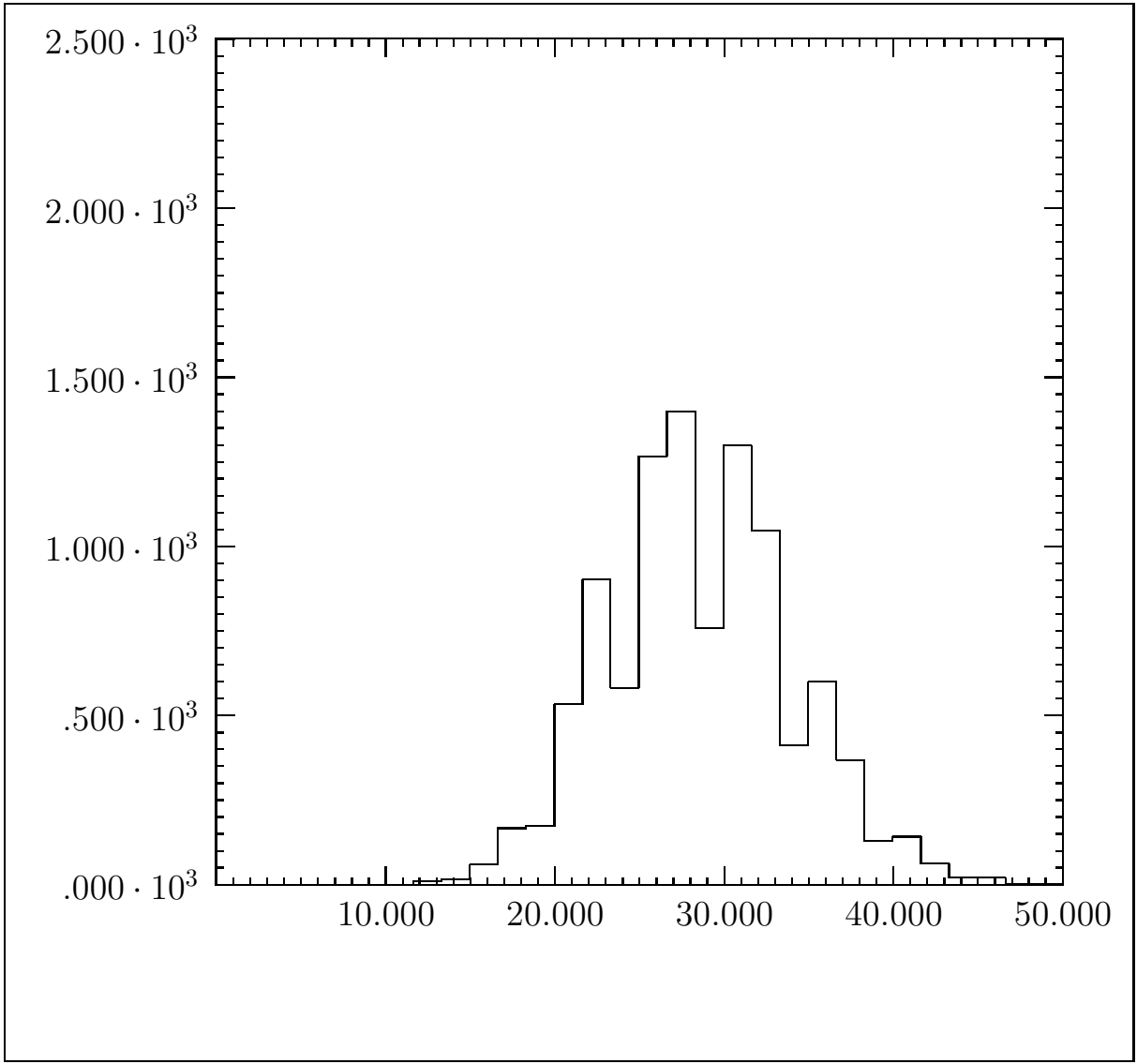
\begin{figure}
\setlength{\unitlength}{0.1mm}
\begin{picture}(1600,1500)
\put(0,0){\framebox(1600,1500){}}
\put(300,250){\begin{picture}(1200,1200)
\put(0,0){\framebox(1200,1200){}}
\multiput(  300.00,0)(  300.00,0){   4}{\line(0,1){25}}
\multiput(     .00,0)(   30.00,0){  41}{\line(0,1){10}}
\multiput(  300.00,1200)(  300.00,0){   4}{\line(0,-1){25}}
\multiput(     .00,1200)(   30.00,  0){  41}{\line(0,-1){10}}
\put( 300,-25){\makebox(0,0)[t]{\large$   2.500\cdot10^{  -1}$}}
\put( 600,-25){\makebox(0,0)[t]{\large$   5.000\cdot10^{  -1}$}}
\put( 900,-25){\makebox(0,0)[t]{\large$   7.500\cdot10^{  -1}$}}
\put(1200,-25){\makebox(0,0)[t]{\large$  10.000\cdot10^{  -1}$}}
\multiput(0,     .00)(0,  240.00){   6}{\line(1,0){25}}
\multiput(0,   24.00)(0,   24.00){  50}{\line(1,0){10}}
\multiput(1200,     .00)(0,  240.00){   6}{\line(-1,0){25}}
\multiput(1200,   24.00)(0,   24.00){  50}{\line(-1,0){10}}
\put(-25,   0){\makebox(0,0)[r]{\large$    .000\cdot10^{   3}$}}
\put(-25, 240){\makebox(0,0)[r]{\large$    .500\cdot10^{   3}$}}
\put(-25, 480){\makebox(0,0)[r]{\large$   1.000\cdot10^{   3}$}}
\put(-25, 720){\makebox(0,0)[r]{\large$   1.500\cdot10^{   3}$}}
\put(-25, 960){\makebox(0,0)[r]{\large$   2.000\cdot10^{   3}$}}
\put(-25,1200){\makebox(0,0)[r]{\large$   2.500\cdot10^{   3}$}}
\end{picture}}
\put(300,250){\begin{picture}(1200,1200)
\thinlines
\newcommand{\x}[3]{\put(#1,#2){\line(1,0){#3}}}
\newcommand{\y}[3]{\put(#1,#2){\line(0,1){#3}}}
\newcommand{\z}[3]{\put(#1,#2){\line(0,-1){#3}}}
\newcommand{\e}[3]{\put(#1,#2){\line(0,1){#3}}}
\y{   0}{   0}{   0}\x{   0}{   0}{  39}
\y{  39}{   0}{   0}\x{  39}{   0}{  40}
\y{  79}{   0}{   1}\x{  79}{   1}{  40}
\y{ 119}{   1}{   0}\x{ 119}{   1}{  40}
\z{ 159}{   1}{   1}\x{ 159}{   0}{  40}
\y{ 199}{   0}{   5}\x{ 199}{   5}{  40}
\y{ 239}{   5}{   0}\x{ 239}{   5}{  40}
\y{ 279}{   5}{   3}\x{ 279}{   8}{  40}
\y{ 319}{   8}{   4}\x{ 319}{  12}{  40}
\z{ 359}{  12}{   3}\x{ 359}{   9}{  40}
\y{ 399}{   9}{   4}\x{ 399}{  13}{  40}
\y{ 439}{  13}{   1}\x{ 439}{  14}{  40}
\y{ 479}{  14}{   1}\x{ 479}{  15}{  40}
\y{ 519}{  15}{   0}\x{ 519}{  15}{  40}
\y{ 559}{  15}{   4}\x{ 559}{  19}{  40}
\y{ 599}{  19}{   7}\x{ 599}{  26}{  40}
\y{ 639}{  26}{  12}\x{ 639}{  38}{  40}
\z{ 679}{  38}{   3}\x{ 679}{  35}{  40}
\y{ 719}{  35}{  13}\x{ 719}{  48}{  40}
\y{ 759}{  48}{  11}\x{ 759}{  59}{  40}
\z{ 799}{  59}{   1}\x{ 799}{  58}{  40}
\y{ 839}{  58}{  18}\x{ 839}{  76}{  40}
\y{ 879}{  76}{  10}\x{ 879}{  86}{  40}
\y{ 919}{  86}{  15}\x{ 919}{ 101}{  40}
\y{ 959}{ 101}{  26}\x{ 959}{ 127}{  40}
\y{ 999}{ 127}{  52}\x{ 999}{ 179}{  40}
\y{1039}{ 179}{  36}\x{1039}{ 215}{  40}
\y{1079}{ 215}{  90}\x{1079}{ 305}{  40}
\y{1119}{ 305}{ 189}\x{1119}{ 494}{  40}
\y{1159}{ 494}{ 706}\x{1159}{1200}{  40}
\end{picture}}
\end{picture}
\vspace{1in}
\caption{
$v$--distribution in $u+u\to u+u+n(G)$ at $\protect\sqrt s=15.4$ TeV.}
\label{fthree}
\end{figure}
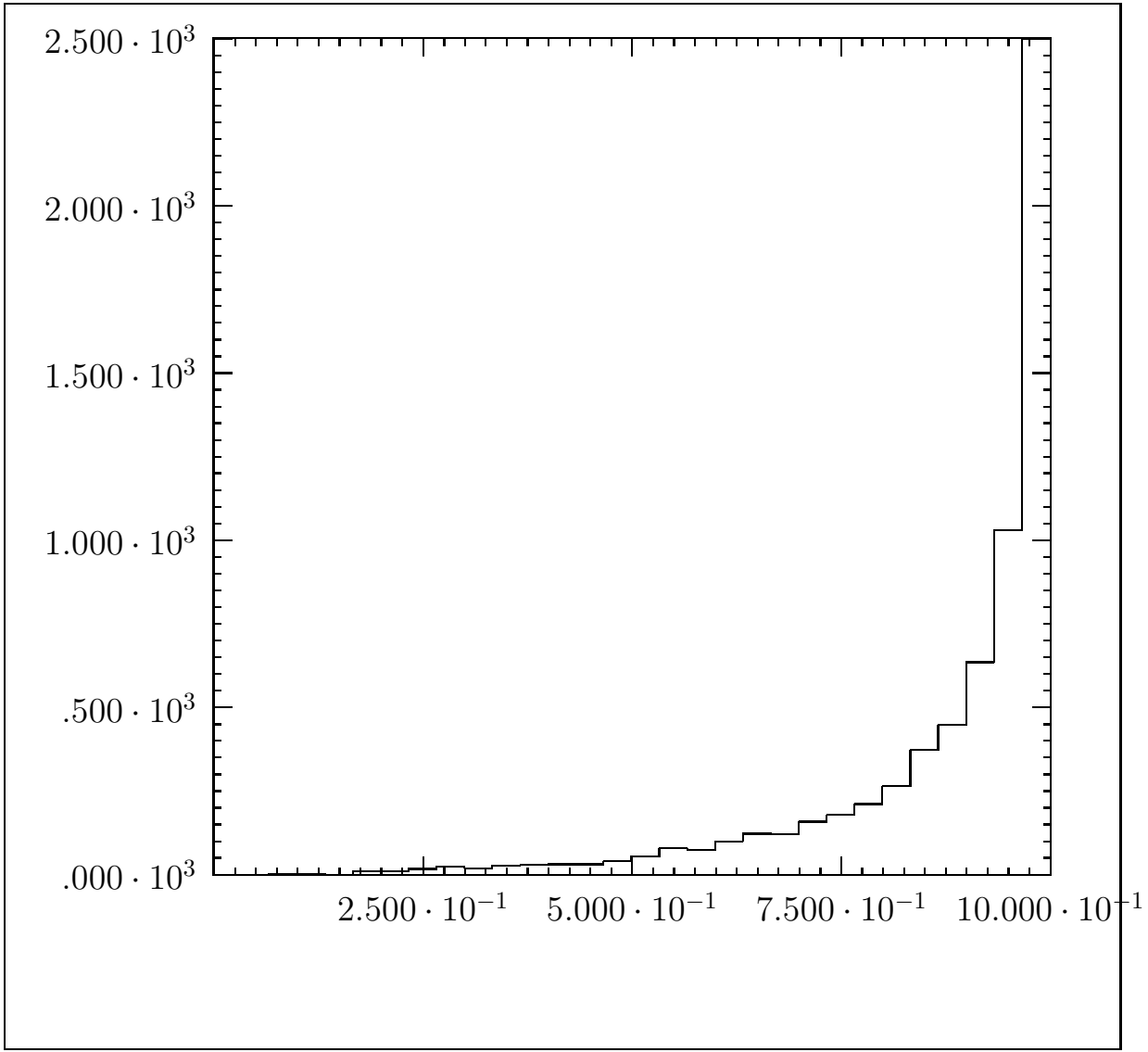
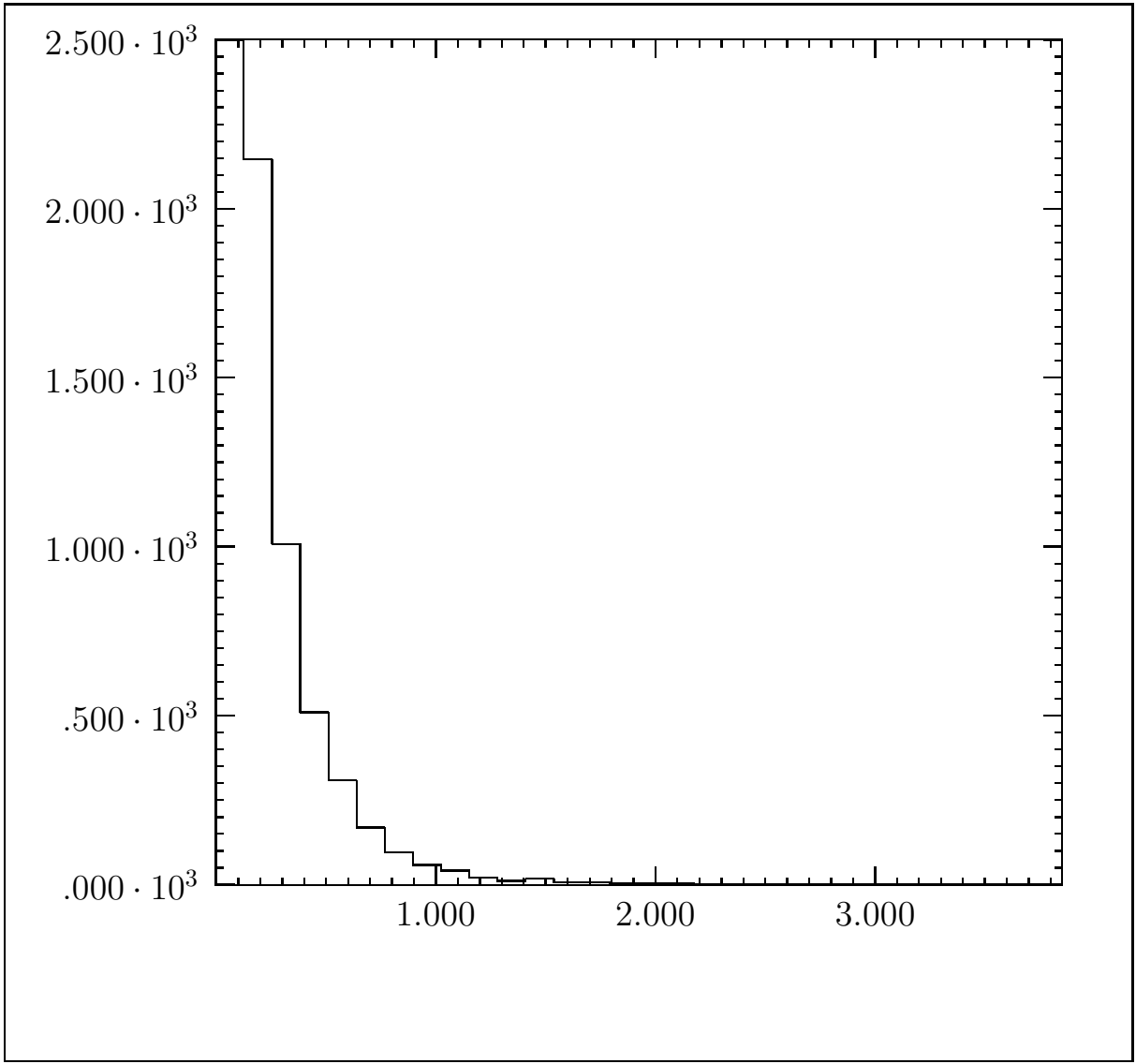
\begin{figure}
\setlength{\unitlength}{0.1mm}
\begin{picture}(1600,1500)
\put(0,0){\framebox(1600,1500){}}
\put(300,250){\begin{picture}(1200,1200)
\put(0,0){\framebox(1200,1200){}}
\multiput(  311.69,0)(  311.69,0){   3}{\line(0,1){25}}
\multiput(     .00,0)(   31.17,0){  39}{\line(0,1){10}}
\multiput(  311.69,1200)(  311.69,0){   3}{\line(0,-1){25}}
\multiput(     .00,1200)(   31.17,  0){  39}{\line(0,-1){10}}
\put( 311,-25){\makebox(0,0)[t]{\large$   1.000$}}
\put( 623,-25){\makebox(0,0)[t]{\large$   2.000$}}
\put( 935,-25){\makebox(0,0)[t]{\large$   3.000$}}
\multiput(0,     .00)(0,  240.00){   6}{\line(1,0){25}}
\multiput(0,   24.00)(0,   24.00){  50}{\line(1,0){10}}
\multiput(1200,     .00)(0,  240.00){   6}{\line(-1,0){25}}
\multiput(1200,   24.00)(0,   24.00){  50}{\line(-1,0){10}}
\put(-25,   0){\makebox(0,0)[r]{\large$    .000\cdot10^{   3}$}}
\put(-25, 240){\makebox(0,0)[r]{\large$    .500\cdot10^{   3}$}}
\put(-25, 480){\makebox(0,0)[r]{\large$   1.000\cdot10^{   3}$}}
\put(-25, 720){\makebox(0,0)[r]{\large$   1.500\cdot10^{   3}$}}
\put(-25, 960){\makebox(0,0)[r]{\large$   2.000\cdot10^{   3}$}}
\put(-25,1200){\makebox(0,0)[r]{\large$   2.500\cdot10^{   3}$}}
\end{picture}}
\put(300,250){\begin{picture}(1200,1200)
\thinlines
\newcommand{\x}[3]{\put(#1,#2){\line(1,0){#3}}}
\newcommand{\y}[3]{\put(#1,#2){\line(0,1){#3}}}
\newcommand{\z}[3]{\put(#1,#2){\line(0,-1){#3}}}
\newcommand{\e}[3]{\put(#1,#2){\line(0,1){#3}}}
\y{   0}{   0}{1200}\x{   0}{1200}{  39}
\z{  39}{1200}{ 170}\x{  39}{1030}{  40}
\z{  79}{1030}{ 546}\x{  79}{ 484}{  40}
\z{ 119}{ 484}{ 239}\x{ 119}{ 245}{  40}
\z{ 159}{ 245}{  97}\x{ 159}{ 148}{  40}
\z{ 199}{ 148}{  67}\x{ 199}{  81}{  40}
\z{ 239}{  81}{  35}\x{ 239}{  46}{  40}
\z{ 279}{  46}{  18}\x{ 279}{  28}{  40}
\z{ 319}{  28}{   8}\x{ 319}{  20}{  40}
\z{ 359}{  20}{  10}\x{ 359}{  10}{  40}
\z{ 399}{  10}{   5}\x{ 399}{   5}{  40}
\y{ 439}{   5}{   3}\x{ 439}{   8}{  40}
\z{ 479}{   8}{   5}\x{ 479}{   3}{  40}
\y{ 519}{   3}{   0}\x{ 519}{   3}{  40}
\z{ 559}{   3}{   1}\x{ 559}{   2}{  40}
\y{ 599}{   2}{   0}\x{ 599}{   2}{  40}
\y{ 639}{   2}{   0}\x{ 639}{   2}{  40}
\z{ 679}{   2}{   2}\x{ 679}{   0}{  40}
\y{ 719}{   0}{   0}\x{ 719}{   0}{  40}
\y{ 759}{   0}{   0}\x{ 759}{   0}{  40}
\y{ 799}{   0}{   0}\x{ 799}{   0}{  40}
\y{ 839}{   0}{   0}\x{ 839}{   0}{  40}
\y{ 879}{   0}{   0}\x{ 879}{   0}{  40}
\y{ 919}{   0}{   0}\x{ 919}{   0}{  40}
\y{ 959}{   0}{   0}\x{ 959}{   0}{  40}
\y{ 999}{   0}{   0}\x{ 999}{   0}{  40}
\y{1039}{   0}{   0}\x{1039}{   0}{  40}
\y{1079}{   0}{   0}\x{1079}{   0}{  40}
\y{1119}{   0}{   0}\x{1119}{   0}{  40}
\y{1159}{   0}{   0}\x{1159}{   0}{  40}
\end{picture}}
\end{picture}
\vspace{1in}
\caption{Total transverse
momentum of gluons in $u+u\to u+u+n(G)$ at $\protect\sqrt s=15.4$~TeV. The
units of the horizontal axis are TeV.}
\label{ffour}
\end{figure}

Our results are shown in Figs.~\ref{ftwo}, \ref{fthree} and \ref{ffour}, where
we show respectively the gluon multiplicity, the distribution of $v=(s-s')/s$,
where $s'$ is the squared invariant mass of the final quark--quark system, and
the total gluon transverse momentum in TeV units.
What we see
is that there is a pronounced effect from the multi--gluon radiation at
$\sqrt s=15.4$~TeV: the average value of the number of
radiated gluons with energy $E_G>3$~GeV is, e.g., at $\sqrt s=15.4$~TeV,
\begin{equation}
\langle n_G\rangle=28.5 \pm 5.5.
\label{enineteen}
\end{equation}
The average value of $v$ is, for $\sqrt s=15.4$~TeV,
\begin{equation}
\langle v\rangle=0.92 \pm 0.14.
\label{etwenty}
\end{equation}
The average value of the total transverse momentum in gluons is%
\begin{equation}
\langle p_{\perp,tot}\rangle\equiv\langle|\sum_{i=1}^n\vec k_{i
\perp}|\rangle= (0.19 \pm 0.23) ~{\rm TeV}.
\label{etwentyone}
\end{equation}
This means that a substantial amount of incoming and outgoing energy is
radiated
into gluons (due to the usual collinearity of radiation with its source, most
of
the final state radiation would be a part of the typical jet associated with
its
parent quark). Entirely similar results hold for the SSC $\sqrt s= 40$~
TeV case. Evidently, any realistic treatment of QCD corrections to
corrections to LHC/SSC
processes must analyze the full event--by--event $n$--gluon effects as they
interact with the detector efficiencies and cuts. Our SSCBHLG event generator
provides the first amplitude--based exponentiated Monte Carlo realization of
such $n$--gluon effects and we hope to participate in a study of their
interplay
with detector effects elsewhere \cite{toapp}.\par
An important consequence of our work is a prediction for the effect on the
overall normalization of an LHC/SSC physics process due to multiple--gluon
radiation, in an unambiguous way, since all IR--singular effects have canceled
out of our calculation. For example, if we simply compute the $O(\alpha_s)$
soft
cross section, $E_G<3$~GeV, we get (for $u+u\to u+u+(G)$, $\sqrt s=40$~TeV)
that
\begin{equation}
\sigma/\sigma_{Born}=1+\delta_{vir}+\delta_{real}=-22.1,
\end{equation}
whereas if we compute the
cross section for $u+u\to u+u+n(G)$ and require $v<3$~GeV$/7.7$~TeV$=.00039$,
we
get
\begin{equation}
\sigma/\sigma_{Born}=1.2\times 10^{-11}.
\label{etwentythree}
\end{equation}
Hence, we see that multi--gluon radiation is
crucial to getting the proper normalization of the cross section.\par
We emphasize that the results for $\sqrt s=40$~TeV give similar conclusions
\cite{toapp} to those in Figs.~\ref{ftwo}--\ref{ffour}. Entirely analogous
results hold for the pure initial--state multigluon event generator SSCYFSG,
except only two lines radiate, so that $\langle n_G\rangle$ is reduced by a
factor $\sim2$, with the appropriate corresponding reductions in $\langle v
\rangle$ and $\langle p_{\perp,tot}\rangle$. Such results are effectively an
integration over the events used in Figs.~\ref{ftwo}--\ref{ffour}, wherein one
integrates over all gluons within some cones close to the outgoing fermions. We
do not show a separate set of plots for SSCYFSG here but we will present a more
detailed discussion of the respective initial--state $n$--gluon radiation
elsewhere \cite{toapp}. Here, we are focusing on the complete gluon radiation
problem.\par
\section{Conclusions}\label{concl}\par
In this paper, we have shown that the IR singularities in fermion--(anti--)%
fermion
scattering at LHC/SSC energies in QCD cancel at order $\alpha_s$, permitting
an immediate extension of YFS QED exponentiation methods to such processes. The
YFS Monte Carlo methods invented by two of us \cite{jdwd} (S.J. and B.F.L.W.)
for QED radiation in $Z^0$ physics can then be extended to multi--gluon Monte
Carlo event generators with well--defined IR behavior to all orders in
$\alpha_s$. Such a Monte Carlo realization of amplitude--based $n$--gluon
effects
opens the way to a new era in higher--order QCD corrections to hard processes
such as those of interest to the LHC/SSC physics program.\par
More specifically, we have realized our exponentiated, n--gluon effects via the
Monte Carlo event generators SSCBHLG and SSCYFSG. Since the latter amounts to
an
initial--state restriction of the former, we have presented results from the
former generator in this paper. We find, for example, that the average number
of
radiated gluons in $u+u\to u+u+n(G)$ at $\sqrt s=15.4$~TeV is $28.5 \pm 5.5$
and
that $\langle v\rangle=0.92 \pm 0.14$. Further, the cross--section
normalization is strongly affected by our $n$--gluon effects. Hence, only by
taking these effects into account, as they interact with detector simulation,
for example, can a truly detailed picture of the SSC/LHC physics platform be
obtained. We hope to participate in such an investigation elsewhere.\par
In summary, we have presented, for the first time, the realistic
event--by--event Monte Carlo realization of multi--gluon effects in hard QCD
processes at SSC/LHC energies in which IR singularities are canceled to
all orders
in $\alpha_s$. A new approach to LHC/SSC physics has thus been created. We
look forward with excitement to its many applications!\par

\newpage
\input feynman
\begin{figure}
\mbox{}
\vspace{-1in}
\begin{center}
\begin{picture}(1600,-1000)
\drawline\gluon[\S\REG](0,1600)[8]
\drawline\fermion[\W\REG](\gluonbackx,\gluonbacky)[4000]
\drawline\fermion[\E\REG](\gluonbackx,\gluonbacky)[4000]
\drawline\fermion[\W\REG](\gluonfrontx,\gluonfronty)[4000]
\drawline\fermion[\E\REG](\gluonfrontx,\gluonfronty)[4000]
\put(0,-9000){\makebox(0,0){\Large\bf(a)}}
\end{picture}
\end{center}
\begin{center}
\begin{picture}(1600,10500)
\drawline\gluon[\S\REG](-19000,-1400)[5]
\drawline\fermion[\W\REG](\gluonbackx,\gluonbacky)[7000]
\drawline\fermion[\E\REG](\gluonbackx,\gluonbacky)[7000]
\drawline\fermion[\NW\REG](\gluonfrontx,\gluonfronty)[4000]
\Xtwo\fermionbackx \Ytwo\fermionbacky
\drawline\fermion[\NE\REG](\gluonfrontx,\gluonfronty)[4000]
\drawline\gluon[\W\REG](\fermionbackx,\fermionbacky)[5]
\drawline\fermion[\E\REG](\fermionbackx,\fermionbacky)[4000]
\drawline\fermion[\W\REG](\Xtwo,\Ytwo)[4000]
\drawline\fermion[\E\REG](\Xtwo,\Ytwo)[250]
\put(-20000,-10000){\makebox(0,0)[l]{\Large(i)}}
\put(-8000,-2000){\makebox(0,0)[l]{\Large+}}
\end{picture}
\begin{picture}(1600,10500)
\drawline\gluon[\S\REG](0,1500)[5]
\drawline\fermion[\W\REG](\gluonfrontx,\gluonfronty)[7000]
\drawline\fermion[\E\REG](\gluonfrontx,\gluonfronty)[7000]
\drawline\fermion[\SW\REG](\gluonbackx,\gluonbacky)[4000]
\Xtwo\fermionbackx \Ytwo\fermionbacky
\drawline\fermion[\E\REG](\Xtwo,\Ytwo)[250]
\drawline\fermion[\SE\REG](\gluonbackx,\gluonbacky)[4000]
\Xfour\fermionbackx \Yfour\fermionbacky
\drawline\gluon[\W\FLIPPED](\Xfour,\Yfour)[5]
\drawline\fermion[\E\REG](\Xfour,\Yfour)[4000]
\drawline\fermion[\W\REG](\Xtwo,\Ytwo)[4000]
\put(0,-10000){\makebox(0,0)[l]{\Large(ii)}}
\put(8000,-2000){\makebox(0,0)[l]{\Large+}}
\end{picture}
\begin{picture}(1600,10500)
\drawline\gluon[\S\REG](18000,1600)[8]
\drawline\fermion[\W\REG](\gluonbackx,\gluonbacky)[4000]
\drawline\fermion[\E\REG](\gluonbackx,\gluonbacky)[4000]
\drawline\fermion[\W\REG](\gluonfrontx,\gluonfronty)[4000]
\drawline\fermion[\E\REG](\gluonfrontx,\gluonfronty)[4000]
\drawline\gluon[\S\REG](22000,1600)[8]
\drawline\fermion[\E\REG](\gluonfrontx,\gluonfronty)[4000]
\drawline\fermion[\E\REG](\gluonbackx,\gluonbacky)[4000]
\put(20000,-10000){\makebox(0,0)[l]{\Large(iii)}}
\end{picture}
\end{center}
\begin{center}
\begin{picture}(1600,13500)
\drawline\fermion[\E\REG](-22000,900)[7000]
\Xtwo\fermionbackx \Ytwo\fermionbacky
\Xthree\fermionfrontx \Ythree\fermionfronty
\drawline\gluon[\SE\REG](\Xthree,\Ythree)[7]
\Xone\gluonbackx \Yone\gluonbacky
\drawline\gluon[\SW\REG](\Xtwo,\Ytwo)[7]
\drawline\fermion[\W\REG](\Xthree,\Ythree)[2000]
\drawline\fermion[\E\REG](\Xtwo,\Ytwo)[2000]
\drawline\fermion[\E\REG](\gluonbackx,\gluonbacky)[8000]
\drawline\fermion[\E\REG](\Xone,\Yone)[2000]
\drawline\fermion[\W\REG](\gluonbackx,\gluonbacky)[2000]
\put(-20000,-9000){\makebox(0,0)[l]{\Large(iv)}}
\put(-12000,-2000){\makebox(0,0)[l]{\Large+}}
\end{picture}
\begin{picture}(1500,13500)
\drawline\gluon[\S\REG](-5000,-2500)[4]
\drawline\fermion[\E\REG](\gluonbackx,\gluonbacky)[6000]
\drawline\fermion[\W\REG](\gluonbackx,\gluonbacky)[6000]
\Xone\gluonfrontx \Yone\gluonfronty
\drawline\gluon[\NE\FLIPPED](\Xone,\Yone)[4]
\Xtwo\gluonbackx \Ytwo\gluonbacky \advance\Ytwo by 50
\drawline\gluon[\NW\REG](\Xone,\Yone)[4]
\drawline\fermion[\E\REG](\Xtwo,\Ytwo)[2000]
\drawline\fermion[\W\REG](\gluonbackx,\gluonbacky)[2000]
\drawline\fermion[\W\REG](\Xtwo,\Ytwo)[10000]
\put(-6600,-9000){\makebox(0,0)[l]{\Large(v)}}
\put(1500,-2000){\makebox(0,0)[l]{\Large+}}
\end{picture}
\begin{picture}(1000,13500)
\drawline\gluon[\S\REG](8600,2100)[4]
\drawline\fermion[\E\REG](\gluonfrontx,\gluonfronty)[6000]
\drawline\fermion[\W\REG](\gluonfrontx,\gluonfronty)[6000]
\Xone\gluonbackx \Yone\gluonbacky
\drawline\gluon[\SW\FLIPPED](\Xone,\Yone)[4]
\drawline\fermion[\W\REG](\gluonbackx,\gluonbacky)[2000]
\drawline\gluon[\SE\REG](\Xone,\Yone)[4]
\drawline\fermion[\E\REG](\gluonbackx,\gluonbacky)[2000]
\drawline\fermion[\W\REG](\gluonbackx,\gluonbacky)[10000]
\put(6600,-9000){\makebox(0,0)[l]{\Large(vi)}}
\put(0,-10000){\makebox(0,0)[l]{\Large\bf(b)}}
\put(16000,-2000){\makebox(0,0)[l]{\Large+}}
\end{picture}
\begin{picture}(1200,13500)
\startphantom
\drawline\gluon[\S\REG](20000,1600)[8]
\stopphantom
\Xone\particlemidx \Yone\particlemidy
\drawline\fermion[\W\REG](\gluonbackx,\gluonbacky)[2500]
\drawline\fermion[\E\REG](\gluonbackx,\gluonbacky)[2500]
\drawline\fermion[\W\REG](\gluonfrontx,\gluonfronty)[2500]
\drawline\fermion[\E\REG](\gluonfrontx,\gluonfronty)[2500]
\drawline\gluon[\N\FLIPPED](\gluonbackx,\gluonbacky)[3]
\advance\gluonbackx by10 \advance\gluonbacky by-10
\drawline\fermion[\N\REG](\gluonbackx,\gluonbacky)[500]
\drawline\gluon[\S\REG](20000,1600)[3]
\drawline\fermion[\S\REG](\gluonbackx,\gluonbacky)[500]
\put(18000,-9000){\makebox(0,0)[l]{\Large(vii)}}
\advance\Xone by-300
\put(\Xone,\Yone){\circle*{2000}}
\put(23000,-2000){\makebox(0,0)[l]{\Large+~f.s.e.}}
\put(24000,-9000){\makebox(0,0)[l]{\Large(viii)}}
\end{picture}
\end{center}
\begin{center}
\begin{picture}(1600,15500)
\drawline\gluon[\S\REG](-20000,1600)[8]
\drawline\fermion[\W\REG](\gluonbackx,\gluonbacky)[4000]
\Xone\particlemidx \Yone\particlemidy
\drawline\fermion[\E\REG](\gluonbackx,\gluonbacky)[4000]
\drawline\fermion[\W\REG](\gluonfrontx,\gluonfronty)[4000]
\drawline\fermion[\E\REG](\gluonfrontx,\gluonfronty)[4000]
\drawline\gluon[\SE\REG](\Xone,\Yone)[2]
\put(-21000,-12000){\makebox(0,0)[l]{\Large(i)}}
\put(-15000,-2000){\makebox(0,0)[l]{\Large+}}
\end{picture}
\begin{picture}(1600,15500)
\drawline\gluon[\S\REG](-10000,1600)[8]
\drawline\fermion[\W\REG](\gluonbackx,\gluonbacky)[4000]
\drawline\fermion[\E\REG](\gluonbackx,\gluonbacky)[4000]
\drawline\fermion[\W\REG](\gluonfrontx,\gluonfronty)[4000]
\Xone\particlemidx \Yone\particlemidy
\drawline\fermion[\E\REG](\gluonfrontx,\gluonfronty)[4000]
\drawline\gluon[\NE\REG](\Xone,\Yone)[2]
\put(-11000,-10000){\makebox(0,0)[l]{\Large(ii)}}
\put(-6000,-2000){\makebox(0,0)[l]{\Large+}}
\end{picture}
\begin{picture}(1600,15500)
\drawline\gluon[\S\REG](0,1600)[8]
\drawline\fermion[\W\REG](\gluonbackx,\gluonbacky)[4000]
\drawline\fermion[\E\REG](\gluonbackx,\gluonbacky)[4000]
\drawline\fermion[\W\REG](\gluonfrontx,\gluonfronty)[4000]
\drawline\fermion[\E\REG](\gluonfrontx,\gluonfronty)[4000]
\drawline\gluon[\NE\REG](\particlemidx,\particlemidy)[2]
\put(-1000,-10000){\makebox(0,0)[l]{\Large(iii)}}
\put(0,-14000){\makebox(0,0)[l]{\Large\bf(c)}}
\put(6000,-2000){\makebox(0,0)[l]{\Large+}}
\end{picture}
\begin{picture}(1600,15500)
\drawline\gluon[\S\REG](10000,1600)[8]
\drawline\fermion[\W\REG](\gluonbackx,\gluonbacky)[4000]
\drawline\fermion[\E\REG](\gluonbackx,\gluonbacky)[4000]
\Xone\particlemidx \Yone\particlemidy
\drawline\fermion[\W\REG](\gluonfrontx,\gluonfronty)[4000]
\drawline\fermion[\E\REG](\gluonfrontx,\gluonfronty)[4000]
\drawline\gluon[\SE\REG](\Xone,\Yone)[2]
\put(9000,-10000){\makebox(0,0)[l]{\Large(iv)}}
\put(15000,-2000){\makebox(0,0)[l]{\Large+}}
\end{picture}
\begin{picture}(1600,15500)
\drawline\gluon[\S\REG](20000,1600)[8]
\Xone\particlemidx \Yone\particlemidy
\drawline\fermion[\W\REG](\gluonbackx,\gluonbacky)[4000]
\drawline\fermion[\E\REG](\gluonbackx,\gluonbacky)[4000]
\drawline\fermion[\W\REG](\gluonfrontx,\gluonfronty)[4000]
\drawline\fermion[\E\REG](\gluonfrontx,\gluonfronty)[4000]
\drawline\gluon[\E\FLIPPEDCENTRAL](\Xone,\Yone)[4]
\put(19000,-10000){\makebox(0,0)[l]{\Large(v)}}
\end{picture}
\end{center}
\vspace{2in}
\caption{The process $q+{
\llap{\protect\phantom q}^{\scriptscriptstyle(}\bar
 q^{\scriptscriptstyle)}{}}'\to q+{\llap{\protect%
\phantom q}^{\scriptscriptstyle(}\bar q^{\scriptscriptstyle)}{}}'+(G)$ to
 $O(\alpha_s)$: (a) Born
approximation; (b) $O(\alpha_s)$ virtual correction; (c) $O(\alpha_s)$
bremsstrahlung process. Here, f.s.e.~represents the fermion self-energies.}
\label{fone}
\end{figure}
%
%
\newpage
\newpage
\newpage
\end{document}